\documentclass[a4paper,12pt]{article}

\usepackage[T2A]{fontenc}
\usepackage[cp1251]{inputenc}
\usepackage[english]{babel}
\usepackage{amssymb,amsfonts,amsmath,mathtext,cite,enumerate,float} 
\usepackage{graphicx}
\usepackage{epstopdf}
\usepackage{hyperref}

\newcommand{\ptl}{\partial}
\newcommand{\co}{\omega_{\rm co}}
\newcommand{\gH}{\mathfrak{H}}
\newcommand{\gT}{\mathfrak{T}}
\newcommand{\gR}{\mathfrak{R}}

\title{SCALAR KLEIN--GORDON EQUATION AND ITS ANALYTICALLY CONTINUED DISPERSION DIAGRAM}

\author{A. I. Korolkov, A. V. Shanin}

\begin{document}

\maketitle
\renewcommand{\abstractname}{\vspace{-\baselineskip}} 

\begin{abstract}	\noindent
The scalar Klein-Gordon equation describes wave motion in a waveguide with a cut-off. For example, the displacement of an elastic cord  anchored to a solid base by elastic elements  can be described by the scalar Klein-Gordon equation.
We analyse this equation  using the concept of analytical continuation of dispersion diagram. Particularly, it is shown that the dispersion diagram is topologically equivalent to a  tube analytically embedded in two-dimensional complex space. The corresponding Fourier integral is studied on this tube using the Cauchy's theorem. The basic properties of the scalar Klein-Gordon equation  are established.

\end{abstract}

\quad\rule{425pt}{0.4pt}

\section{Introduction}
Multilayer acoustical waveguides find its application in geology \cite{BenMenahem2012}, oceanography \cite{Pekeris1948}, and medical physics \cite{Muller2005}.  Such waveguides can be modeled by the matrix Klein-Gordon equation also known as waveguide finite element method \cite{Aalami1973}. The latter can be considered as a set of scalar Klein--Gordon systems interacting with each other. The scalar equation is, thus, an elementary ``building block'' of the theory of multilayer systems. Below we thoroughly study the scalar Klein-Gordon equation using the  analytical continuation of dispersion diagram.

\section{Scalar Klein--Gordon equation}
\label{subsec0421}

Consider a scalar Klein--Gordon equation
\begin{equation}
(c^2 \ptl_x^2 - \ptl_t^2 - \co^2) u(t,x) =\delta(x) \delta(t).
\label{eq07030203}
\end{equation} 
Parameter $c$ has the sense of the limiting velocity in the waveguide, and 
$\co$ is the cut-off frequency. 

\begin{figure}[ht]
\centering
\includegraphics[width=10cm]{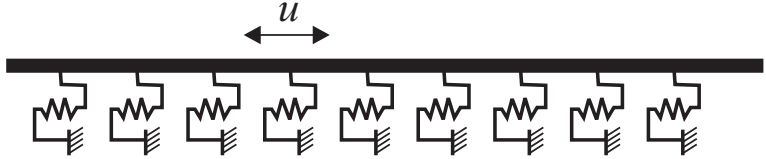}
\caption{Physical model for Klein--Gordon equation. Cord and elastic elements}
\label{figB4201}
\end{figure}

The scalar Klein--Gordon equation 
describes a simple physical model. One can consider the 
variable $u(t,x)$ as longitudinal displacement of an elastic massive cord anchored to 
a solid base by elastic elements (see Fig.~\ref{figB4201}). 
The velocity $c$ is the wave velocity in the cord taken 
without additional elastic elements.
The frequency $\co$
is the resonance frequency of the oscillation of non-deformed cord (taken as a mass) 
on the external springs.   
The source is applied at $x = 0$. 

The solution can be obtained in a standard way. Introduce the Fourier transforms
in the $t$ and $x$ domains:
\begin{equation}
\mathcal{F}_t [s](\omega) \equiv 
\frac{1}{2\pi} \int \limits_{-\infty}^{\infty}
s(t) e^{i \omega t} dt ,
\qquad 
\mathcal{F}_t^{-1} [S](t) \equiv 
\int \limits_{-\infty}^{\infty}
S(\omega) e^{- i \omega t} d\omega ,
\label{eqB13002}
\end{equation} 
\begin{equation}
\mathcal{F}_x [s](k) \equiv 
\frac{1}{2\pi} \int \limits_{-\infty}^{\infty}
s(t) e^{-i k x} dx ,
\qquad 
\mathcal{F}_x^{-1} [S](x) \equiv 
\int \limits_{-\infty}^{\infty}
S(k) e^{i k x} dk .
\label{eqB13003}
\end{equation} 
For the Fourier image $U(\omega, k)$ of $u(t, x)$, get 
\[
(-c^2 k^2 + \omega^2 - \co^2) U(\omega, k) = \frac{1}{4 \pi^2} .
\]
The inverse Fourier transforms in $k$ and $t$ yields
the double integral representation of the field: 
\begin{equation}
u(t,x) = \frac{1}{4\pi^2} 
\int \limits_{-\infty + i\epsilon}^{\infty + i \epsilon}
\int \limits_{-\infty }^{\infty}
\frac{e^{i k x - i \omega t}}{\omega^2 - c^2 k^2 - \co^2} dk \, d\omega.
\label{eq07030204}
\end{equation}
The contours of integration  are chosen according to the causality principle. 

The denominator of (\ref{eq07030204}) is referred to as the dispersion function. 
Its zeros defined by the 
dispersion equation  
\begin{equation}
D(\omega , k) = \omega^2 - c^2 k^2 -\co^2 =0
\label{eqB42001}
\end{equation}  
correspond to modes in the waveguide. 

The dispersion 
equation (\ref{eqB42001}) has order $n = 1$ with respect to $K = k^2$ and  
$W = \omega^2$.
The solution of this equation is given by $\pm k(\omega)$, where
\begin{equation}
k (\omega) = c^{-1} \sqrt{\omega^2 - \co^2}.
\label{eqB42002}
\end{equation}
As the value of $\sqrt{\cdot}$,
we select the branch of the square root that is close to 
$i \co$ when 
$\omega$ is close to zero. 
One can see that for all $\omega$ with ${\rm Im}[\omega] = \epsilon$
the imaginary part of $k(\omega)$ is positive.

\begin{figure}[ht]
\centering
\includegraphics[width=10.39cm]{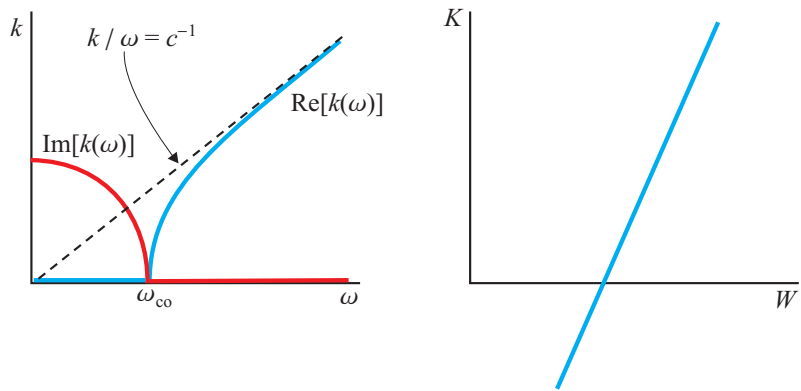}
\caption{Dispersion diagram for a scalar Klein--Gordon equation}
\label{figB0110}
\end{figure}

The graphs of the dispersion diagram in the coordinates $(\omega, k)$ 
and $(W= \omega^2 , K = k^2)$ are shown in Fig.~\ref{figB0110}.  

\section{The field and its analysis}
\label{subsec0422}

Start from the double--integral representation (\ref{eq07030204}).  
Take $x > 0$. Close the contour of integration with respect to $k$ 
in the upper half-plane. For each $\omega$ the integrand has two poles, 
$k = \pm k(\omega)$, and only the pole $k = k(\omega)$ belongs to the 
upper half-plane. Thus, 
by applying the residue theorem we obtain 
\begin{equation}
u(t,x) = - \frac{i}{4\pi c} 
\int \limits_{-\infty + i\epsilon}^{\infty + i \epsilon}
\frac{e^{i k(\omega) x - i \omega t}}{\sqrt{\omega^2 - \co^2}}  d\omega .
\label{eq07030205}
\end{equation}
The integral can be easily taken: 
\begin{equation}
u(t,x) = 
\left\{ 
\begin{array}{ll}
-(2c)^{-1} J_0 (\co \sqrt{t^2 - x^2/c^2}),  & t > x/c, \\
0, & 0 < t < x/c,
\end{array}
\right. 
\label{eq07030206}
\end{equation}
where $J_0$ is the Bessel function.

The following features of the solution should be mentioned: 

\begin{itemize}

\item
The front of the wave propagates with the speed equal to $c$. 
Near the  
front, the system behaves similarly to the usual wave equation. Only the cord 
shown in Fig.~\ref{figB4201} plays role there, not the additional elastic elements.
One can see that the shape of the front is the Heaviside function. The field before the 
front, i.~e.\ for $t < x/c$, is equal to zero.

\item
For $t \gg x/c$ the system displays oscillations in time having 
the frequency tending to the cut-off~$\co$. This means that 
far behind the front there exist oscillations with a small wavenumber and
produced mainly by the elastic elements. 

\end{itemize}

These features agree with the common understanding of the dispersion 
diagram (Fig.~\ref{figB0110}). 
The group velocity depends on $\omega$ as  
\begin{equation}
v_{\rm gr} = \left( 
\frac{dk}{d\omega}
\right)^{-1} = 
c \frac{\sqrt{\omega^2 - \co^2}}{\omega} , 
\qquad 
\co < \omega < \infty.
\label{eqB42016a}
\end{equation}
One can see that the group velocity grows from $0$ to $c$ as 
$\omega$ grows from $\co$ to~$\infty$. 

One can roughly imagine the wave process in the Klein--Gordon system as follows. First, 
the fast wave in the cord carries the energy over the waveguide. Next, the energy manifests 
itself in almost in-phase oscillations with frequency~$\co$.


\section{Complex parametrization of the dispersion diagram}
\label{subsec0423}

\begin{figure}[ht]
\centering
\includegraphics[width=10.39cm]{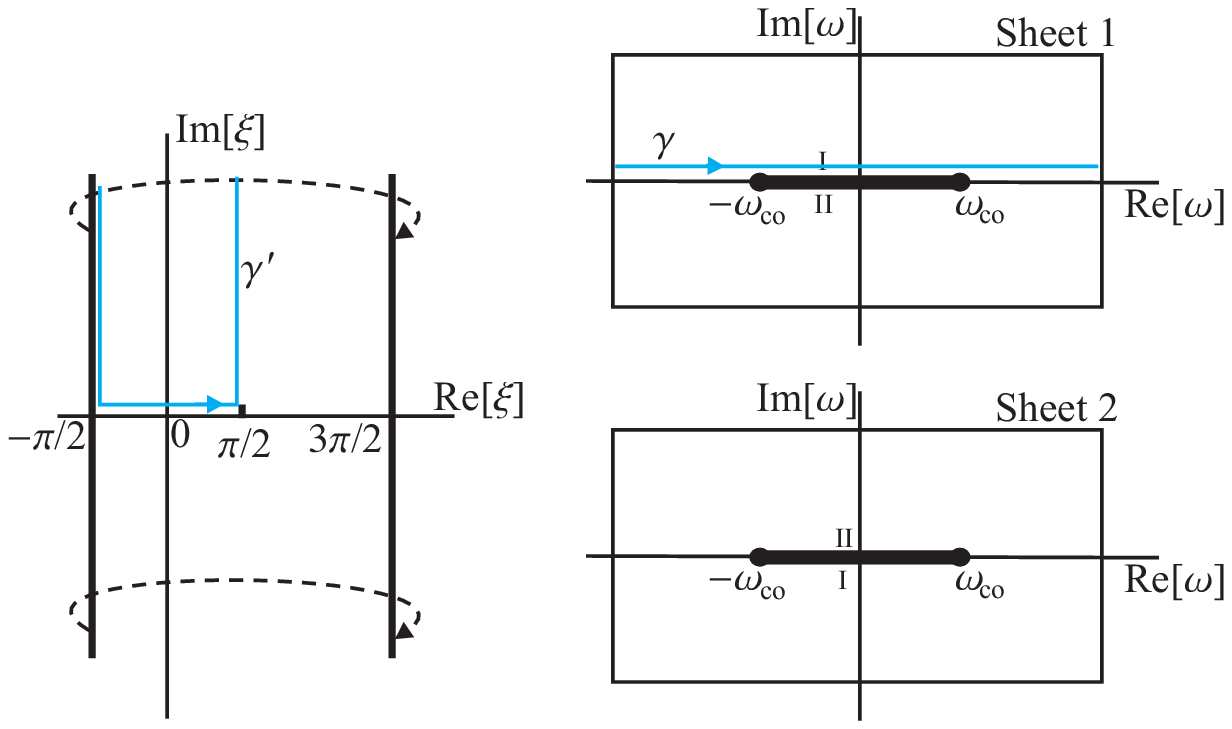}
\caption{Tube $\gT$ of the variable $\xi$ (left). The Riemann 
surface $\gR$ (right)}
\label{figB4202}
\end{figure}

Let us study the dispersion equation 
(\ref{eqB42001}) for complex $\omega$ and~$k$. 
The set of solutions of 
(\ref{eqB42001}) is an analytic manifold $\gH$ of complex dimension~1
in the space $(\omega, k) \in \mathbb{C}^2$.
Thus, $\gH$  
has real dimension~2 in the space of real 
dimension~4.  
Let us build a parametrization of $\gH$ by a complex variable $\xi$ without 
using branching functions (i.~e.\ square roots). This parametrization 
is as follows:
\begin{equation}
\omega = \co \sin \xi, 
\qquad
k = i \frac{\co}{c} \cos \xi,
\label{eqB42003}
\end{equation}
i.~e.\ for each complex $\xi$ the point $(\omega , k)$ defined by 
(\ref{eqB42003}) obeys  (\ref{eqB42001}), and for each solution 
of (\ref{eqB42001}) there exists a corresponding~$\xi$.
One can see that the trigonometric functions in (\ref{eqB42003}) are periodic with  
(real) period $2 \pi$. Thus, it would be natural to consider $\xi$ being defined on 
a {\em tube} $\gT$, i.~e.\ on a strip  
$-\pi/2 \le {\rm Re}[\xi] \le 3\pi/2$ with the edges attached to each other
(see Fig.~\ref{figB4202}, left). 

Formulae (\ref{eqB42003}) establish a bijection between $\gT$ and~$\gH$. This means that the complex dispersion diagram  $\gH$ is topologically a tube.    

Manifold $\gH$ is embedded in $\mathbb{C}^2$ analytically since the derivatives 
$d \omega / d\xi$ and $d k / d \xi$ are not equal to zero 
simultaneously.
Variable $\xi$ can be used as a local variable  
on $\gH$ everywhere (in the context of analytic manifolds).

The Riemann surface $\gR$ of the function $k(\omega)$ defined by (\ref{eqB42002})  
has two sheets cut along the segment $[-\co , \co]$. The points 
$\pm \co$ are branch points of the surface, both having second order.
The scheme of this Riemann surface is shown in Fig.~\ref{figB4202}, right.  
In the scheme, we mark  by 
equal Roman numbers the shores of the cut attached to each other.   Indeed, 
$\gR$ topologically is a tube as well. 

Consider the representation (\ref{eq07030205}). Let us rewrite it in the following form: 
\begin{equation}
u(t, x) =  
- \frac{ i }{4 \pi c^2} 
\int_\gamma
\frac{e^{i k (\omega) - i \omega t}}{k(\omega)} d \omega. 
\label{eq0703025a}
\end{equation}
Contour $\gamma$ is $(-\infty + i \epsilon , \infty + i \epsilon)$
passed in the positive direction. 
Using the new variable $\xi$, 
the integral (\ref{eq0703025a}) can be rewritten using variable~$\xi$:
\begin{equation}
u(t , x) =
 -
\frac{1}{4\pi c} \int_{\gamma'}
e^{i k (\xi) x - i \omega(\xi) t} d\xi,
\label{eqB42012d}
\end{equation}
where $k(\xi)$ and $\omega(\xi)$ are defined by (\ref{eqB42003}), 
and the contour $\gamma'$ is shown in Fig.~\ref{figB4202}. 
 
Thus, the integration is performed on $\gT$ or on $\gH$.
We can say that the representation 
of the field can be interpreted as an integral over some contour drawn on the 
dispersion diagram~$\gH$. 
This statement remains true when a general waveguide is considered. We remind that
 one can define contour 
integration on any analytic manifold of complex dimension~1 (see~\cite{Shabat1992}), 
and the Cauchy's theorem remains valid, i.~e.\ one can deform the integration 
contour on the manifold. Indeed, the differential 1-form that is integrated over the 
contour should be holomorphic, which is generally the case. 


\section{Deformation of the integration contour}
\label{subsec0424}

The contour of integration $\gamma'$ can be deformed freely in the 
finite part of $\gT$. We also study admissible manipulations with the 
``tails'' of $\gamma'$ going to infinity, i.~e.\ we study the possibilities of closing the contour. 

The first sheet of $\gR$ corresponds to the domain ${\rm Im}[\xi] > 0$ 
of $\gT$. Respectively, the second sheet corresponds to 
${\rm Im}[\xi] < 0$. The cut corresponds to the real segment 
$-\pi / 2 \le \xi \le 3\pi / 2$ is cutting $\gT$ into two parts.  

There are two outlets of the tube, namely ${\rm Im}[\xi] \to  \infty$ and 
${\rm Im}[\xi] \to  -\infty$. These outlets are ``infinities'', in other words,
$|\omega| \to \infty$ there.   
  
\begin{figure}[ht]
\centering
\includegraphics[width=10.39cm]{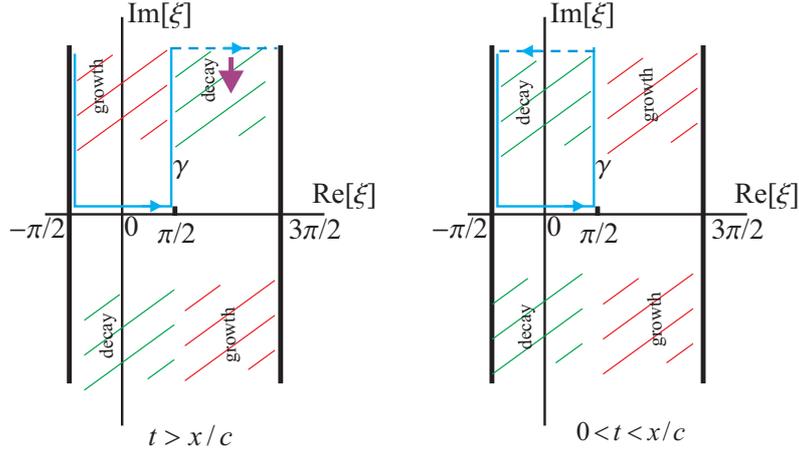}
\caption{Domains of exponential growth and decay of the integrand of (\ref{eq07030205})
for $t > x/c$ (left), and for $0<t<x/c$ (right)}
\label{figB4203}
\end{figure}

Let be $t > x/ c$.    
Consider the exponential factor (the integrand of (\ref{eqB42012d})) 
on the tube~$\gT$. 
The domains of exponential growth and decay of the exponential factor in this case
are shown in Fig.~\ref{figB4203}, left.
One can see that 
\begin{equation}
\exp \{ i k(\omega) - i \omega t  \} =
\exp \left \{
i \co \left( 
\frac{ix}{c} \cos \xi - t \sin \xi 
\right)
\right \} =
\exp \{ i \co r \cos (\xi - \eta) \},
\label{eqB42004}
\end{equation}   
where $r$ and $\eta$ are coordinates linked to $t$ and $x$ by
\begin{equation}
\frac{ix }{c} = r \cos\eta ,
\qquad 
- t = r \sin \eta.
\label{eqB42005}
\end{equation}
The inverse coordinate change is
\begin{equation}
r = \sqrt{t^2 - x^2 / c^2},
\qquad
\eta = {\rm atan} (i c t / x).
\label{eqB42005a}
\end{equation}
Parameter $r$ is positive real, and parameter $\eta$ takes values on the 
half-line 
\[
\eta \in (-\pi/2  , -\pi / 2 + i \infty). 
\]

As we mentioned above, the Cauchy's theorem is valid on $\gT$, thus the 
integration contour can be deformed on $\gT$. Moreover, one can add some 
remote segments/arcs to the contour $\gamma'$ in the domain of decay of the exponential factor, or, the same, to deform the tails of $\gamma'$ in the domain of 
exponential decay.
  
Add an infinitely remote segment shown by the dashed line in Fig.~\ref{figB4203}, left.
As the result, contour $\gamma'$ becomes a closed loop going around the tube.
It can be transformed into the real segment $[-\pi/2 , 3\pi/2]$.

If $0< t < x/c$, the consideration can be made again, and the domains of  growth and decay 
become having form shown in Fig.~\ref{figB4203}, right. The way to close the contour is shown in the figure by a dashed line. One can see that  the contour encircles a domain having no singularities, and thus the integral is equal to zero. 

\begin{figure}[ht]
\centering
\includegraphics[width=6cm]{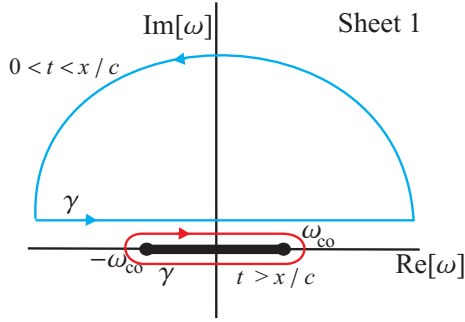}
\caption{Closings of the integration contour on $\gR$
for $t > x/ c$ (red) and for $0 < t < x/c$ (blue)}
\label{figB4204}
\end{figure}

The consideration above has been made in terms of  
$\gT$ and the ``natural'' variable $\xi$. The same consideration can be translated 
into the language of the variable $\omega$ and the Riemann surface~$\gR$.
The closings of the integration contour 
$\gamma$ in the cases 
$t > x/ c$ and $0 < t < x/c$
are shown in Fig.~\ref{figB4204}. Indeed, the consideration 
in the variable $\omega$ is completely equivalent to the consideration in the variable~$\xi$.

We can make the following statements: 

\begin{itemize}

\item
The fact that the field before the front ($0 <t < x/ c$) 
is equal to zero has a topological nature. 
The field is equal to zero because $\gamma'$ can be closed in such
a way  that it becomes 
homological to a zero contour. 

\item  
After the front ($t > x/c$) the contour encircles the ``neck'' of the tube. 
According to the Cauchy's theorem, one can take any of such contours. Below we choose a contour that passes through the saddle points of the integrand of (\ref{eqB42012d}). 

\end{itemize}

 
\section{``Far--field'' asymptotics of the solution}
\label{subsec0425}

Let be $t > x/c$.
Compute asymptotic estimations of the field directly from (\ref{eq0703025a}), i.~e.\
without using the solution (\ref{eq07030206}). 
First, perform the formal saddle-point analysis. 
Introduce the the parameter $V = x / t$ 
and assume that $x \to \infty$ while $V$ is fixed. 

The integrand of (\ref{eq0703025a}) contains the exponential factor  
\[
\exp \{ i k(\omega) x - i \omega t\} = \exp \{ i x (k(\omega) - V^{-1} \omega  ) \}. 
\]
For building an asymptotic representation for $u(t, x) = u(x/ V , x)$, 
one should find the stationary points of the function 
$k(\omega) - V^{-1} \omega$. 
We are looking for the  saddle points 
$\omega_*$ such that 
\begin{equation}
\frac{d k(\omega_*)}{d\omega}
= 
V^{-1}. 
\label{eqB42013}
\end{equation}
Note that this condition is equivalent to 
$v_{\rm gr} = V$.

According to (\ref{eqB42002}), the saddle--points are $\pm \omega_*$,
\begin{equation}
\omega_* = \co \left( 1 - V^2 / c^2 \right)^{-1/2}.
\label{eqB42014}
\end{equation}
In the variable $\xi$, the saddle--points corresponding to $\pm \omega_*$
are $\xi_*$ and $\xi_* - \pi$, where $\xi_* = \eta + \pi$.

\begin{figure}[ht]
\centering
\includegraphics[width=10.39cm]{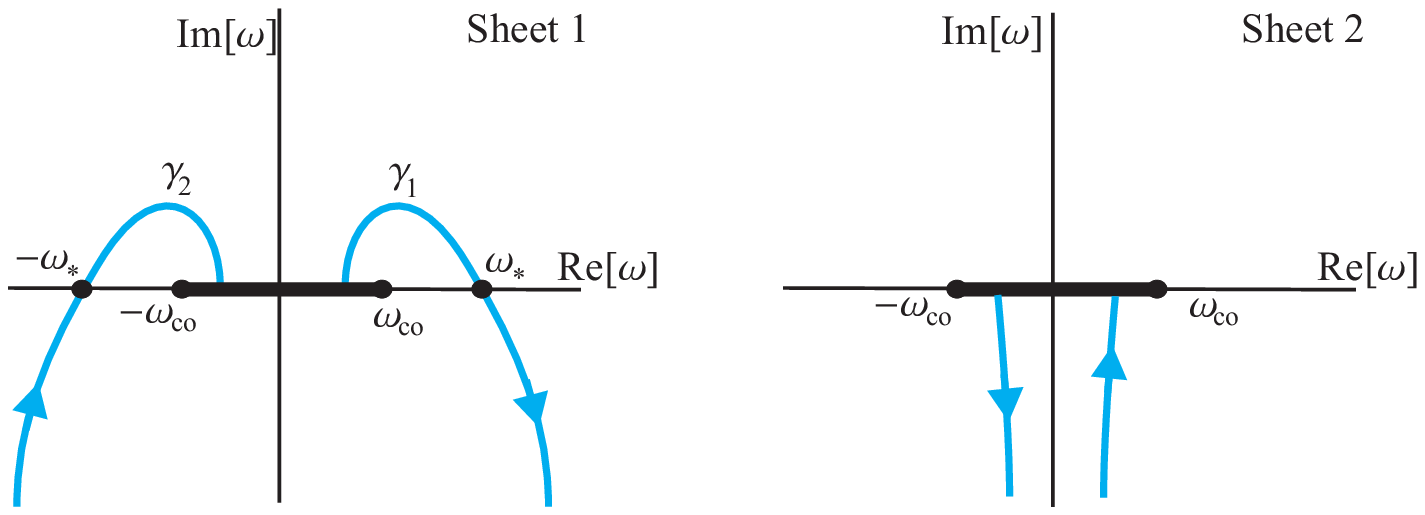}
\caption{Steepest descend contours on $\gR$}
\label{figB4208}
\end{figure}

Let us build
the steepest descend contours passing 
through the saddle-points. As it is known, they are the contours on which the value 
${\rm Re}[k(\omega) - V^{-1} \omega]$ is constant. 
The sketch of these 
contours is shown in Fig.~\ref{figB4208}. 
One can see that
contour $\gamma$ can be transformed into the sum of the 
steepest descend   
contours $\gamma_1$ and $\gamma_2$ passing through $\omega_*$ and $-\omega_*$,
respectively. 

%
%
%

The integral can be estimated by the saddle--point expressions.  The representations 
corresponding to the points $\omega_*$ and $-\omega_*$ are
\begin{equation}
u_{\pm \omega_*} (t, \omega) \approx 
-\frac{1}{2c} 
\frac{\exp \{ \mp i \co \sqrt{t^2 - x^2 / c^2} \pm i\pi / 4 \}}{
\sqrt{2\pi \co} \left( t^2 - x^2 / c^2 \right)^{1/4}}. 
\label{eqB42015}
\end{equation}
The sum of these terms provide, indeed, a
large-argument  asymptotics for 
(\ref{eq07030206}).
This asymptotics is valid for a large argument of Bessel function, i.~e.\ for 
\begin{equation}
\omega_{\rm co} \sqrt{t^2 - x^2 / c^2} \gg 1. 
\label{eqB42021z}
\end{equation}


\section{``Near--field'' asymptotics of the solution}

Consider the integral (\ref{eqB42012d}).
Our aim is to obtain the near-field asymptotics. 
Conversely to the 
concept of the far--field, where the exponential factor should be 
rapidly oscillating, here we 
are trying
to find a contour integration (homotopic to $\gamma'$) 
on which the exponential factor does not change considerably.
In this case the integrand can be substituted by a constant.  

\begin{figure}[ht]
\centering
\includegraphics[width=10.39cm]{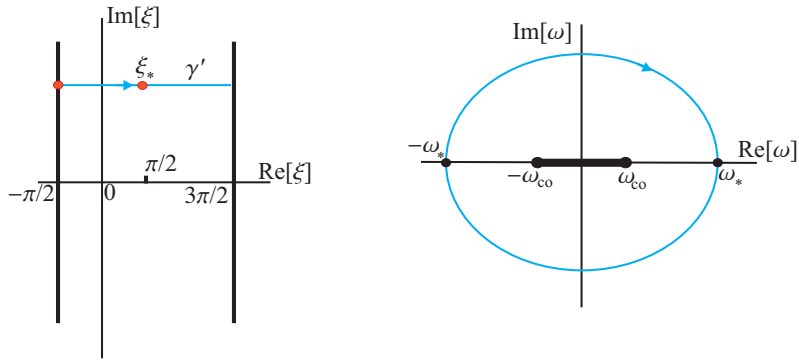}
\caption{Integration contour for getting the near-field asymptotics (\ref{eqB42021b})}
\label{figB4208a}
\end{figure}

Consider the contour $\gamma'$ that is the
segment 
$\gamma' = [\eta  , \eta + 2\pi ]$, or, the same 
$\gamma' = [\xi_* - \pi  , \xi_* + \pi ]$. 
According to formula (\ref{eqB42004}), the exponential term 
changes on this contour weakly if $r \omega_{\rm co} \ll 1$ or  
 if
\begin{equation}
\omega_{\rm co} \sqrt{t^2 - x^2 / c^2} \ll 1. 
\label{eqB42021a}
\end{equation}
The contour $\gamma'$ in the $\xi$-domain is shown in Fig.~\ref{figB4208a}, left. 
The image of this contour in the $\omega$-domain 
is shown in Fig.~\ref{figB4208a}, right. The contour passes through the points 
$\eta = \xi_* - \pi$ and $\xi_*$. In the $\omega$-plane their images 
are the saddle--points
$-\omega_*$, $\omega_*$ defined by (\ref{eqB42014}). Note if the point 
$(t,x)$ is close to the wave front (i.~e.\ if $t - x/ c$ is small), 
the contour  of integration in the 
$\omega$-domain is a large ellipse.

Under this condition, a simple estimation 
of the integral (\ref{eqB42012d})
can be obtained. 
Since the exponential factor changes slightly, the exponential factor can be estimated   
by the constant $\exp\{i \omega_{\rm co} r \cos \eta \}$, 
and this constant is close to~1.
Thus,(\ref{eqB42012d}) can be estimated as
\begin{equation}
u(t, x) \approx - \frac{1}{2c},
\label{eqB42021b}
\end{equation}
which is, indeed, an asymptotics of (\ref{eq07030206}) for small values of the 
argument of Bessel function.   
The condition of validity (\ref{eqB42021a}) makes sense: the argument of Bessel function in 
(\ref{eq07030205}) is close to zero in this case, and the value of Bessel function is close to~1.



\section{Acknowledgements}
The work is supported by the RFBR grant 19-29-06048.

\bibliographystyle{unsrt}
\bibliography{references_kor}

\end{document}